\documentclass[aps,prd,nofootinbib,amsmath,amssymb,superscriptaddress,showpacs,showkeys,twocolumn,superscriptaddress,10pt]{revtex4}%superscriptaddress,showpacs,showkeys,

\pdfoutput=1

\usepackage{graphicx}
\usepackage{dcolumn}
\usepackage{bm}
\usepackage{amssymb}
\usepackage{latexsym}
\usepackage{booktabs}
\usepackage{amsmath}
\usepackage{multirow}
\usepackage[colorlinks=true, linkcolor=red, citecolor=blue]{hyperref}

\newcommand{\be}{\begin{equation}}
\newcommand{\ee}{\end{equation}}
\newcommand{\bq}{\begin{eqnarray}}
\newcommand{\eq}{\end{eqnarray}}

\usepackage[usenames,dvipsnames]{xcolor}

\DeclareMathAlphabet\mathbfcal{OMS}{cmsy}{b}{n}
\definecolor{darkgreen}{cmyk}{0.85,0.2,1.00,0.2}
\definecolor{purple}{cmyk}{0.5,1.0,0,0}

\def\barray{\begin{array}}
\def\earray{\end{array}}
\def\be{\begin{equation}}
\def\ee{\end{equation}}
\def\ben{\begin{equation} \nonumber}
\def\een{\end{equation}}
\def\ban{\begin{eqnarray*}}
\def\ean{\end{eqnarray*}}
\def\ba{\begin{eqnarray}}
\def\ea{\end{eqnarray}}

\def\({\left(}
\def\){\right)}

\begin{document}

\title{Models of vacuum energy interacting with cold dark matter: Constraints and comparison}
%\date{}                                           % Activate to display a given date or no date
\author{Hai-Li Li}
\affiliation{Department of Physics, College of Sciences,
Northeastern University, Shenyang 110819, China}
%\email{1329750467@qq.com}
\author{Lu Feng}
\affiliation{College of Physical Science and Technology, Shenyang Normal University, Shenyang
110034, China}
\author{Jing-Fei Zhang}
\affiliation{Department of Physics, College of Sciences,
Northeastern University, Shenyang 110819, China}
\author{Xin Zhang\footnote{Corresponding author}}
\email{zhangxin@mail.neu.edu.cn}
\affiliation{Department of Physics, College of Sciences,
Northeastern University, Shenyang 110819, China}
\affiliation{Ministry of Education Key Laboratory of Data Analytics and Optimization
for Smart Industry, Northeastern University, Shenyang 110819, China}
\affiliation{Center for High Energy Physics, Peking University, Beijing 100080, China}

\begin{abstract}

In this paper, we investigate the observational constraints on the scenario of vacuum energy interacting with cold dark matter. We consider eight typical interaction forms in such an interacting vacuum energy scenario. The observational data used in this work to constrain these models include the JLA sample of type Ia supernovae observation, the Planck 2015 distance priors data of cosmic microwave background anisotropies observation, the baryon acoustic oscillations data, and the Hubble constant direct measurement. We find that the current observational data almost equally favor these interacting vacuum energy models. We also find that for all these models of vacuum energy interacting with cold dark matter the case of no interaction is actually well consistent with the current observational data within 1$\sigma$ range.

\end{abstract}

\maketitle
\section{Introduction}\label{sec:intro}

The acceleration of the expansion of the universe was discovered by the observations of type Ia supernovae \cite{Riess:1998cb,Perlmutter:1998np}, and subsequently was confirmed by the observations of cosmic microwave background and large scale structure \cite{Spergel:2003cb,Bennett:2003bz,Tegmark:2003ud,Abazajian:2004aja}. The cosmic acceleration is one of the most important topics studied in modern cosmology. To explain the accelerated expansion of the universe, an exotic form of energy with negative pressure, called ``dark energy'', within the framework of general relativity, has been proposed (for reviews, see, e.g., Refs. \cite{Sahni:2006pa,Bamba:2012cp,Weinberg:1988cp,Peebles:2002gy,Copeland:2006wr,Frieman:2008sn,Sahni:2008zz,Li:2011sd,Kamionkowski:2007wv}). Dark energy makes up about 68\% of the total energy density, and thus the evolution of the current universe is dominated by the dark energy.

The cosmological constant $\Lambda$, first proposed by Einstein in 1917, is known to be the simplest dark energy candidate, with the equation-of-state parameter being $w_{\Lambda} \equiv p_{\Lambda}/\rho_{\Lambda}=-1$. The cosmological model with $\Lambda$ and cold dark matter (CDM) is usually called the $\Lambda$CDM model, which is considered to be the standard model of cosmology. The $\Lambda$CDM model is in excellent agreement with current cosmological observations \cite{Ade:2015xua} and its parameters have been determined to an impressive accuracy by the current observational data. However, the cosmological constant $\Lambda$ has always been plagued with some well-known theoretical difficulties, such as the ``fine-tuning'' and ``cosmic coincidence'' problems \cite{Sahni:1999gb,Bean:2005ru}. %In addition, it is hard to believe that the 6-parameter base $\Lambda$CDM model can completely describe the whole evolution history of the universe, and
%Thus it is necessary to extend the base $\Lambda$CDM model,
Thus, some extensions to the base $\Lambda$CDM cosmology have been widely considered, among which the scenario of vacuum energy interacting with CDM has attracted lots of attention \cite{Amendola:1999er,Amendola:1999qq,TocchiniValentini:2001ty,Amendola:2001rc,Comelli:2003cv,Chimento:2003iea,Cai:2004dk,Zhang:2004gc,Ferrer:2004nv,Zimdahl:2005bk,Zhang:2005rj,Wang:2006qw,Sadjadi:2006qp,Barrow:2006hia,Sasaki:2006kq,Abdalla:2007rd,Bean:2007ny,Guo:2007zk,Bertolami:2007zm,Boehmer:2008av,He:2008tn,CalderaCabral:2008bx,Bean:2008ac,Szydlowski:2008by,Chen:2008ft,Valiviita:2008iv,Couderc:2009tq,Chimento:2009hj,CalderaCabral:2009ja,Majerotto:2009np,Valiviita:2009nu,He:2009mz,He:2009pd,Koyama:2009gd,Li:2009zs,Xia:2009zzb,Cai:2009ht,He:2010ta,Cui:2010dr,Li:2010eu,Gavela:2010tm,Martinelli:2010rt,He:2010im,Chen:2011rz,Fu:2011ab,Clemson:2011an,Li:2011ga,Xu:2011tsa,Zhang:2012uu,Xu:2013jma,Zhang:2013zyn,Wang:2013qy,Salvatelli:2013wra,Yang:2014gza,yang:2014vza,Wang:2014oga,Faraoni:2014vra,Yin:2015pqa,Fan:2015rha,Cai:2015zoa,Duniya:2015nva,Feng:2016djj,Murgia:2016ccp,Sola:2016jky,Sola:2016ecz,Sola:2016zeg,Pourtsidou:2016ico,Costa:2016tpb,Xia:2016vnp,vandeBruck:2016hpz,Kumar:2016zpg,Kumar:2017dnp,Santos:2017bqm,Sola:2017jbl,Guo:2017hea,Zhang:2017ize,Feng:2018yew,Yang:2018euj,Guo:2018ans,Zhao:2018fjj,Feng:2019mym,Li:2019loh}. Following the terminology of Refs.~\cite{Li:2015vla,Guo:2017deu,Guo:2018gyo,Feng:2017usu}, in this paper such a scenario is called the I$\Lambda$CDM cosmology.

When some direct, non-gravitational interaction between vacuum energy and CDM is considered, the continuity equations for the vacuum energy and the CDM can be written as $\dot{\rho}_{\rm vac}=-Q$ and $\dot{\rho}_{\rm c}+3H\rho_{\rm c}=Q$, respectively, where $Q$ denotes the phenomenological interaction term describing the energy transfer rate between vacuum energy and dark matter due to the interaction \cite{Zhang:2005rg,Zhang:2007uh,Zhang:2009qa,Li:2010ak,Zhang:2013lea,Li:2013bya,Li:2014eha,Geng:2015ara,Li:2014cee}. Here, $Q>0$ denotes vacuum energy decaying into dark matter, $Q<0$ denotes dark matter decaying into vacuum energy, and $Q=0$ denotes no interaction between vacuum energy and dark matter. Usually, the form of $Q$ is assumed to be proportional to the density of dark sectors \cite{Amendola:1999qq,Billyard:2000bh}, i.e., $Q\propto \rho$ with $\rho$ here being the density of vacuum energy or cold dark matter or some combination of them. The form of $Q=\beta H \rho$ has often been considered in the literature, of which the benefit is that its numerical calculation is rather convenient. Another choice is the form of $Q=\beta H_{0} \rho$, for which someone argues that the benefit is the form being free of the global expansion (note that it is argued that local interaction should not be relevant to the global expansion). In the above expressions, the dimensionless parameter $\beta$ denotes the coupling strength.

Different phenomenological models of I$\Lambda$CDM can be built by constructing different forms of $Q$. In this work, we will collect the popular forms of $Q$ in the current literature and make a comprehensive analysis for the I$\Lambda$CDM models from the perspective of observational constraints. We wish to investigate which concrete model is more favored by the current observational data. In this work, we consider the following eight typical forms of $Q$: $Q_1=\beta H_{0}\rho_{\rm{vac}}$ \cite{Clemson:2011an}, $Q_2=\beta H_{0}\rho_{\rm c}$ \cite{Boehmer:2008av,Valiviita:2008iv}, $Q_3=\beta H_{0}(\rho_{\rm{vac}}+\rho_{\rm c})$ \cite{Li:2017usw}, $Q_4=\beta H_{0}\frac{\rho_{\rm vac}\rho_{\rm c}}{\rho_{\rm vac}+\rho_{\rm c}}$ \cite{Li:2017usw}, $Q_5=\beta H\rho_{\rm{vac}}$ \cite{Pavon:2007gt}, $Q_6=\beta H\rho_{\rm c}$ \cite{Amendola:2006dg,Guo:2007zk}, $Q_7=\beta H(\rho_{\rm{vac}}+\rho_{\rm c})$ \cite{delCampo:2005tr,Olivares:2005tb,Wang:2005jx}, and $Q_8=\beta H\frac{\rho_{\rm vac}\rho_{\rm c}}{\rho_{\rm vac}+\rho_{\rm c}}$ \cite{Zhang:2004gc}. In the following, for convenience, we denote the I$\Lambda$CDM models as I$\Lambda$CDM1---I$\Lambda$CDM8 according to the forms of $Q$ given by $Q_1$---$Q_8$.

We will constrain these eight I$\Lambda$CDM models by using the current observational data and then make a comparison for them. We wish to see whether some hints of the existence of nonzero interaction for the I$\Lambda$CDM scenario can be found by this exploration.

%\begin{equation}\label{H_{0}1}
%Q=\beta H_{0}\rho_{\rm{de}},
%\end{equation}
%\begin{equation}\label{H_{0}2}
%Q=\beta H_{0}\rho_{\rm c},
%\end{equation}
%\begin{equation}\label{H_{0}3}
%Q=\beta H_{0}(\rho_{\rm{de}}+\rho_{\rm c}),
%\end{equation}
%\begin{equation}\label{H_{0}4}
%Q=\beta H_{0}\frac{\rho_{\rm de}\rho_{\rm c}}{\rho_{\rm de}+\rho_{\rm c}},
%\end{equation}
%\begin{equation}\label{H1}
%Q=\beta H\rho_{\rm{de}},
%\end{equation}
%\begin{equation}\label{H2}
%Q=\beta H\rho_{\rm c},
%\end{equation}
%\begin{equation}\label{H3}
%Q=\beta H(\rho_{\rm{de}}+\rho_{\rm c}),
%\end{equation}
%\begin{equation}\label{H4}
%Q=\beta H\frac{\rho_{\rm de}\rho_{\rm c}}{\rho_{\rm de}+\rho_{\rm c}},
%\end{equation}

The structure of this paper is organized as follows. In Sec.~\ref{sec2}, we introduce the analysis method and present the observational data used in this paper. In Sec.~\ref{sec3}, we report the constraint results and make some relevant discussions for them. Conclusion is given in Sec.~\ref{sec4}.

\section{Method and data}\label{sec2}

Considering a flat universe, the I$\Lambda$CDM models have three free parameters (for describing the late-time evolution of the universe), i.e., $h$, $\Omega_{\rm m0}$, and $\beta$. In this work, we employ the Markov-chain Monte Carlo (MCMC) package {\tt CosmoMC} to infer the posterior distributions of parameters. We use the current observational data to constrain the models and obtain the best-fit values and the 1--2$\sigma$ confidence level ranges for the parameters.

%We employ the $\chi^2$ statistic to estimate the model parameters. For calculating each data set, we give the specific form of the $\chi^2$ function
%\begin{equation}\label{3.1}
%\chi^{2}_{\xi}=\frac{(\xi_{\rm th}-\xi_{\rm obs})^{2}}{\sigma^{2}_{\xi}}.
%\end{equation}
%For a physical quantity $\xi$, $\xi_{\rm th}$ is the theoretically predicted value, $\xi_{\rm obs}$ is the experimentally measured value of it, and $\sigma_{\xi}$ is the standard deviation.
%The total $\chi^2$ is the sum of all $\chi^2_\xi$,
%\begin{equation}\label{3.2}
%\chi^2=\sum\limits_{\xi} \chi^2_\xi.
%\end{equation}

In this work, we consider several observational data sets, including the JLA compilation of type Ia supernova (SN) data, the cosmic microwave background (CMB) data from the Planck 2015 mission, the baryon acoustic oscillation (BAO) measurements, and the direct measurement of the Hubble constant $H_{0}$. So the total $\chi^2$ function is written as
$ \chi^2=\chi^2_{\rm SN}+\chi^2_{\rm CMB}+\chi^2_{\rm BAO}+\chi^2_{H_0}$.

%\begin{equation}\label{3.3}
%  \chi^2=\chi^2_{\rm SN}+\chi^2_{\rm CMB}+\chi^2_{\rm BAO}+\chi^2_{H_0}.
%\end{equation}

All these I$\Lambda$CDM models have the same parameter number, so we can make a fair comparison for them only by comparing their $\chi^2$ values. However, we also consider the $\Lambda$CDM model ($Q=0$, the parameter number is one less than the I$\Lambda$CDM models) in this paper. Thus, we need to consider the effect of the number of parameters and the data points. We employ the Akaike information criterion (AIC) \cite{AIC1974} and the Bayesian information criterion (BIC) \cite{BIC1978} to do the analysis. The AIC is defined as ${\rm AIC}=-2\ln{\mathcal{L}_{\rm{max}}}+2k$ and ${\rm BIC}=-2\ln{\mathcal{L}_{\rm{max}}}+k\ln{N}$, where $\mathcal{L}_{\rm{max}}$ is the maximum likelihood, $k$ is the number of parameters, and $N$ is the number of data points. In our work, we choose the $\Lambda$CDM as a reference model, and thus for these I$\Lambda$CDM models, we are more concerned with the relative value between them and the $\Lambda$CDM model. So we only need to calculate $\Delta {\rm AIC}=\Delta\chi^2_{\rm{min}}+2\Delta k$ and $\Delta{\rm BIC}=\Delta\chi^2_{\rm{min}}+\Delta k\ln N$.

The observational data we use in this work are listed as follows.

$\bullet$ The SN data: For the type Ia supernovae observations, we employ the Joint-Light-curve Analysis (JLA) data compilation, which consists of 740 type Ia supernovae data points. The redshift range of these observations is $z \in$ [0.01,1.30]. The data include some low-redshift samples (from the SDSS and SNLS) and a few high-redshift samples (from the HST).

$\bullet$ The CMB data: Dark energy could affect the CMB through the comoving angular diameter distance (at the decoupling epoch $z \simeq 1100$) and the late integrated Sachs-Wolfe (ISW) effect. But we cannot accurately measure the late ISW effect at present, so the important information from CMB data for constraining dark energy is the angular diameter distance information. In this paper, we mainly focus on the smooth dark energy, and thus the CMB distance priors can provide the necessary information. So we only use the ``Planck distance priors" from the Planck 2015 data \cite{Ade:2015rim}, including the shift parameter $R$, the ``acoustic scale" $\ell_{\rm{A}}$, and the baryon density $\omega_{\rm{b}}$.

$\bullet$ The BAO data: The BAO distance scale data can be used to break the geometric degeneracy. We adopt four BAO points from the six-degree-field galaxy survey (6dFGS) at $z_{\rm eff}=0.106$ \cite{Beutler:2011hx}, the SDSS main galaxy sample (MGS) at $z_{\rm eff}=0.15$ \cite{Ross:2014qpa}, the baryon oscillation spectroscopic survey (BOSS) ``LOWZ" at $z_{\rm eff}=0.32$ \cite{Anderson:2013zyy}, and the BOSS CMASS at $z_{\rm eff}=0.57$ \cite{Anderson:2013zyy}.

$\bullet$ The $H_{0}$ data: For the Hubble constant direct measurement, we use the value given by Efstathiou \cite{Efstathiou:2013via}, i.e., $H_0=70.6\pm3.3$ km s$^{-1}$ Mpc$^{-1}$, which is a re-analysis of the Cepheid data of Riess et al \cite{Riess:2011yx}.

\section{Results and discussion}\label{sec3}

\begin{table*}[!htp]\small
\setlength\tabcolsep{8pt}
\caption{Summary of the information criteria results.}
\label{table1}
\centering
\renewcommand{\arraystretch}{1.5}
\begin{tabular}{lccc}
\\
\hline
Model  & $\chi^2_{\rm min}$ & $\Delta$AIC & $\Delta$BIC \\
  \hline

$\Lambda$CDM       & $699.3776$
                   & $0$
                   & $0$
                   \\

I$\Lambda$CDM1                & $699.1004$
                   & $1.7228$
                   & $6.3402$
                   \\

I$\Lambda$CDM2              & $699.0236$
                   & $1.6460$
                   & $6.2634$
                   \\

I$\Lambda$CDM3              & $699.0638$
                   & $1.6862$
                   & $6.3036$
                   \\

I$\Lambda$CDM4              & $699.0656$
                   & $1.6880$
                   & $6.3054$
                   \\

I$\Lambda$CDM5              & $699.1012$
                   & $1.7236$
                   & $6.3410$
                   \\

I$\Lambda$CDM6              & $698.5856$
                   & $1.2080$
                   & $5.8254$
                   \\

I$\Lambda$CDM7                     & $698.7272$
                   & $1.3496$
                   & $5.9670$
                   \\

I$\Lambda$CDM8              & $699.0852$
                   & $1.7076$
                   & $6.3250$
                   \\
\hline
\end{tabular}
\end{table*}

\begin{table*}[!htp]\small
\setlength\tabcolsep{5pt}
\caption{Fitting results of the $\Lambda$CDM model and the I$\Lambda$CDM models with $Q=\beta H_{0}\rho$. Best-fit values with $\pm1\sigma$ errors are presented. }
\label{table2}
\renewcommand{\arraystretch}{2}\centering
\begin{tabular}{lccccc}
\\
\hline
Parameter  & $\Lambda$CDM & I$\Lambda$CDM1 & I$\Lambda$CDM2 & I$\Lambda$CDM3 & I$\Lambda$CDM4 \\
  \hline
$\Omega_{\rm{m0}}$ & $0.3236^{+0.0074}_{-0.0080}$
                   & $0.3203^{+0.0072}_{-0.0082}$
                   & $0.3189^{+0.0089}_{-0.0066}$
                   & $0.3201^{+0.0077}_{-0.0080}$
                   & $0.3197^{+0.0077}_{-0.0074}$

                   \\

$\Omega_{\rm{b0}}$ & $0.0498^{+0.0007}_{-0.0007}$
                   & $0.0498^{+0.0008}_{-0.0009}$
                   & $0.0495^{+0.0009}_{-0.0006}$
                   & $0.0498^{+0.0007}_{-0.0009}$
                   & $0.0498^{+0.0008}_{-0.0008}$
                   \\

$\beta$            &...
                   & $-0.0110^{+0.0186}_{-0.0179}$
                   & $-0.0098^{+0.0152}_{-0.0151}$
                   & $-0.0063^{+0.0091}_{-0.0070}$
                   & $-0.0270^{+0.0429}_{-0.0385}$
                   \\

$h$                & $0.6673^{+0.0059}_{-0.0051}$
                   & $0.6691^{+0.0061}_{-0.0054}$
                   & $0.6706^{+0.0048}_{-0.0061}$
                   & $0.6696^{+0.0060}_{-0.0055}$
                   & $0.6696^{+0.0056}_{-0.0056}$
                   \\
\hline
\end{tabular}
%\caption{Fitting results of the models.}
\end{table*}

\begin{table*}[!htp]\small
\setlength\tabcolsep{5pt}
\caption{Fitting results of the $\Lambda$CDM model and the I$\Lambda$CDM models with $Q=\beta H\rho$. Best-fit values with $\pm1\sigma$ errors are presented. }
\label{table3}
\renewcommand{\arraystretch}{2}\centering
\begin{tabular}{lccccc}
\\
\hline
Parameter  & $\Lambda$CDM & I$\Lambda$CDM5& I$\Lambda$CDM6&I$\Lambda$CDM7& I$\Lambda$CDM8\\
  \hline
$\Omega_{\rm{m0}}$ & $0.3236^{+0.0074}_{-0.0080}$
                   & $0.3205^{+0.0072}_{-0.0084}$
                   & $0.3197^{+0.0086}_{-0.0072}$
                   & $0.3204^{+0.0073}_{-0.0080}$
                   & $0.3201^{+0.0076}_{-0.0081}$
                   \\

$\Omega_{\rm{b0}}$ & $0.0498^{+0.0007}_{-0.0007}$
                   & $0.0498^{+0.0008}_{-0.0009}$
                   & $0.0484^{+0.0013}_{-0.0009}$
                   & $0.0490^{+0.0009}_{-0.0008}$
                   & $0.0498^{+0.0008}_{-0.0009}$
                   \\

$\beta$            &...
                   & $-0.0091^{+0.0151}_{-0.0157}$
                   & $-0.0064^{+0.0066}_{-0.0045}$
                   & $-0.0038^{+0.0044}_{-0.0039}$
                   & $-0.0202^{+0.0323}_{-0.0309}$
                   \\

$h$                & $0.6673^{+0.0059}_{-0.0051}$
                   & $0.6692^{+0.0064}_{-0.0052}$
                   & $0.6796^{+0.0079}_{-0.0112}$
                   & $0.6748^{+0.0074}_{-0.0071}$
                   & $0.6697^{+0.0058}_{-0.0057}$
                   \\
\hline
\end{tabular}
%\caption{Fitting results of the models.}
\end{table*}

\begin{figure*}[!htp]
\includegraphics[width=13cm]{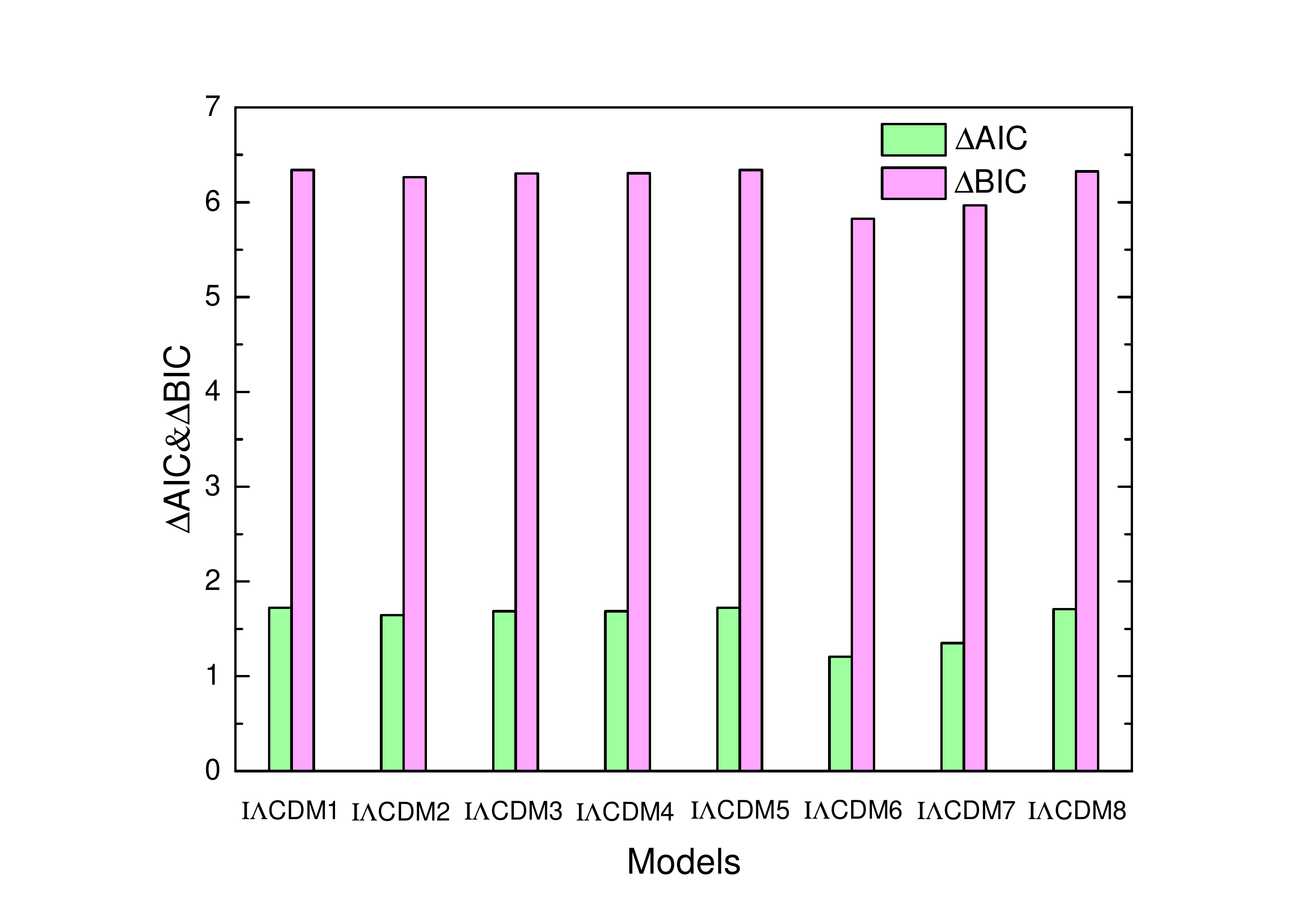}
\caption{\label{fig1}Graphical representation of the results of $\Delta$AIC and $\Delta$BIC for the I$\Lambda$CDM models.}
\end{figure*}

\begin{figure*}[!htp]
\includegraphics[width=15cm]{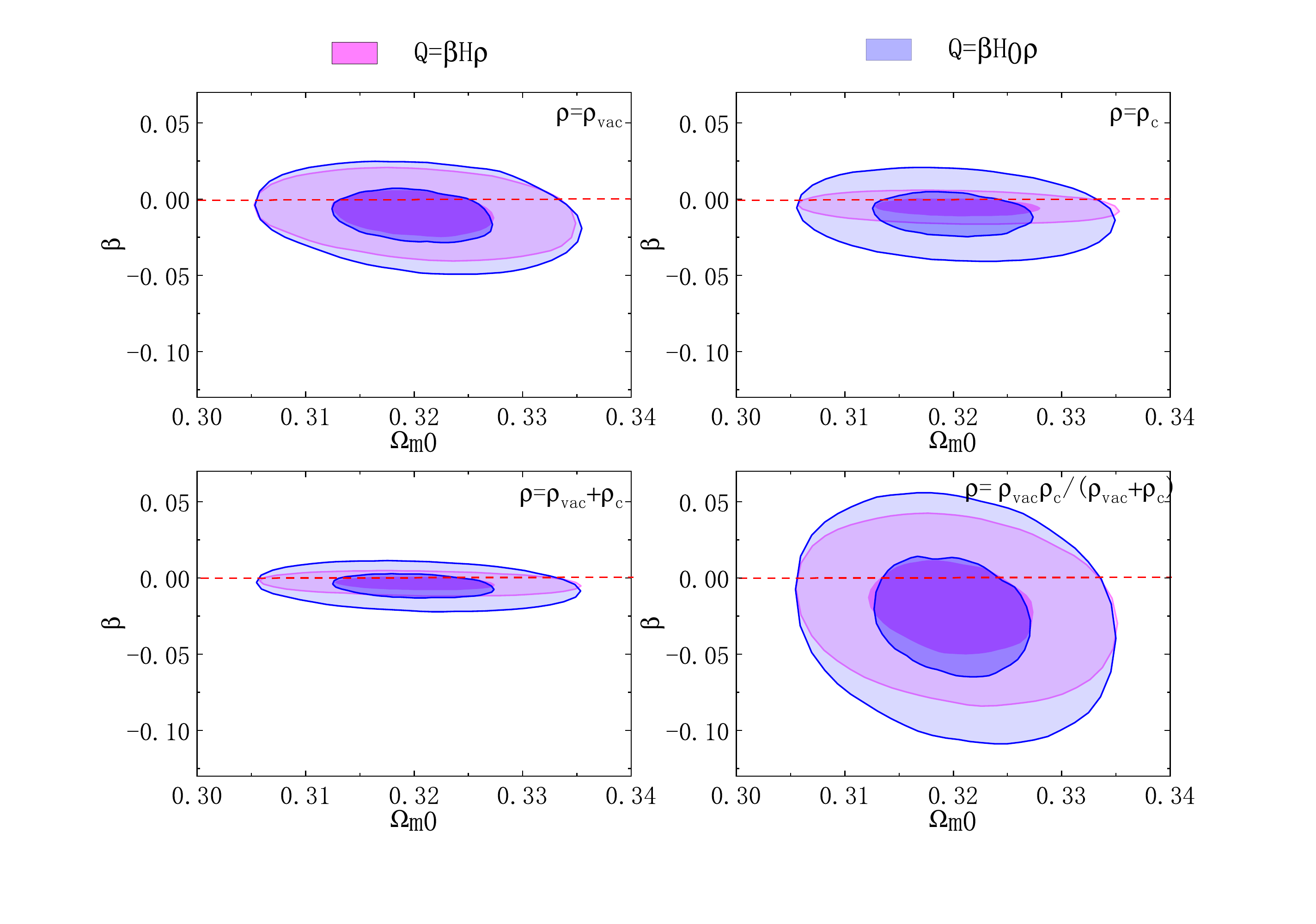}
\caption{\label{fig2}The SN+CMB+BAO+$H_0$ constraints on the I$\Lambda$CDM models. The 68.3\% and 95.4\% confidence level contours are shown in the $\Omega_{\rm{m0}}$--$\beta$ plane. The red dashed line denotes the case of $\beta=0$. }
\end{figure*}

\begin{figure*}[!htp]
\includegraphics[width=15cm]{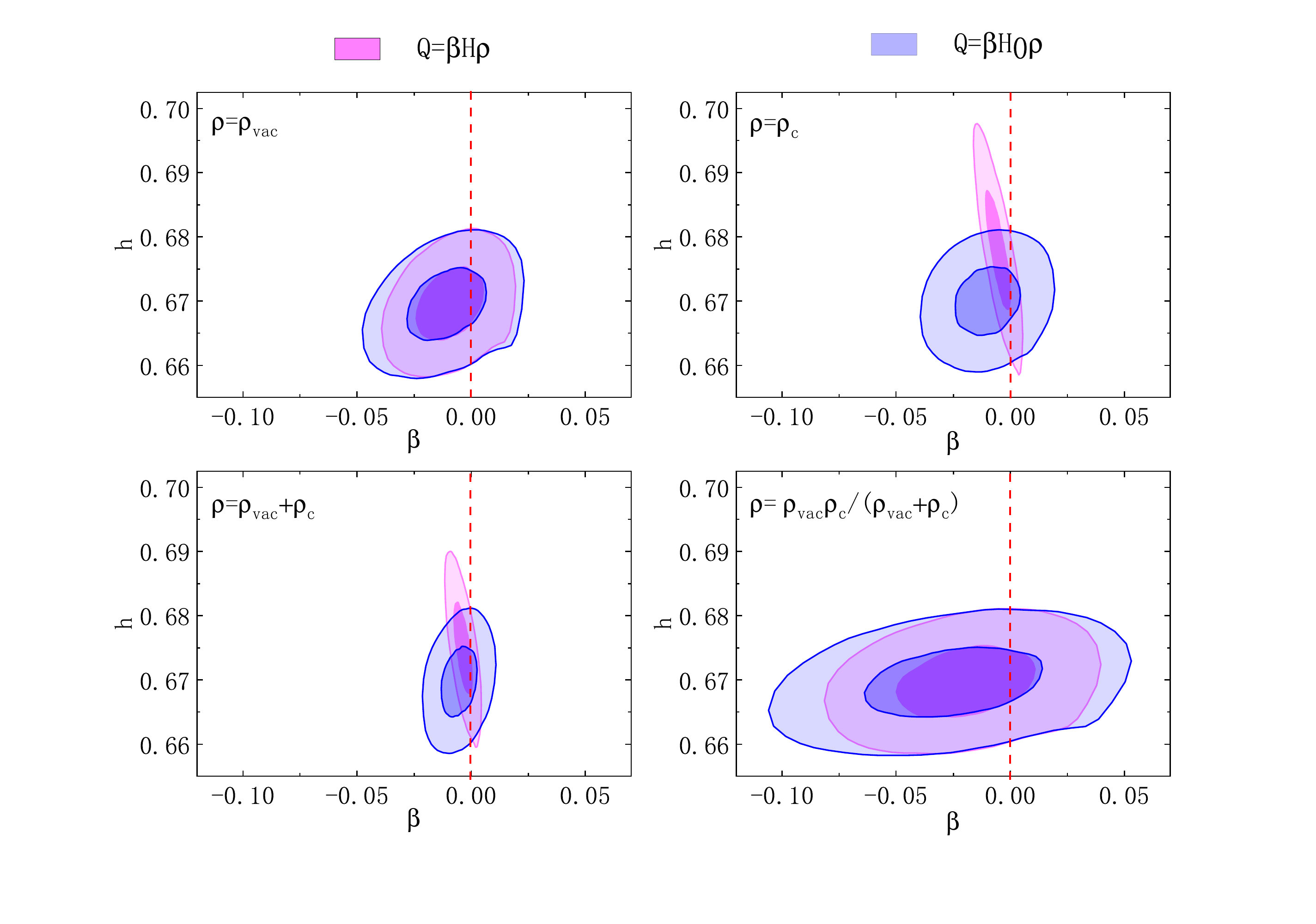}
\caption{\label{fig3}The SN+CMB+BAO+$H_0$ constraints on the I$\Lambda$CDM models. The 68.3\% and 95.4\% confidence level contours are shown in the $\beta$--$h$ plane. The red dashed line denotes the case of $\beta=0$. }
\end{figure*}

\begin{figure*}[!htp]
\includegraphics[width=13cm]{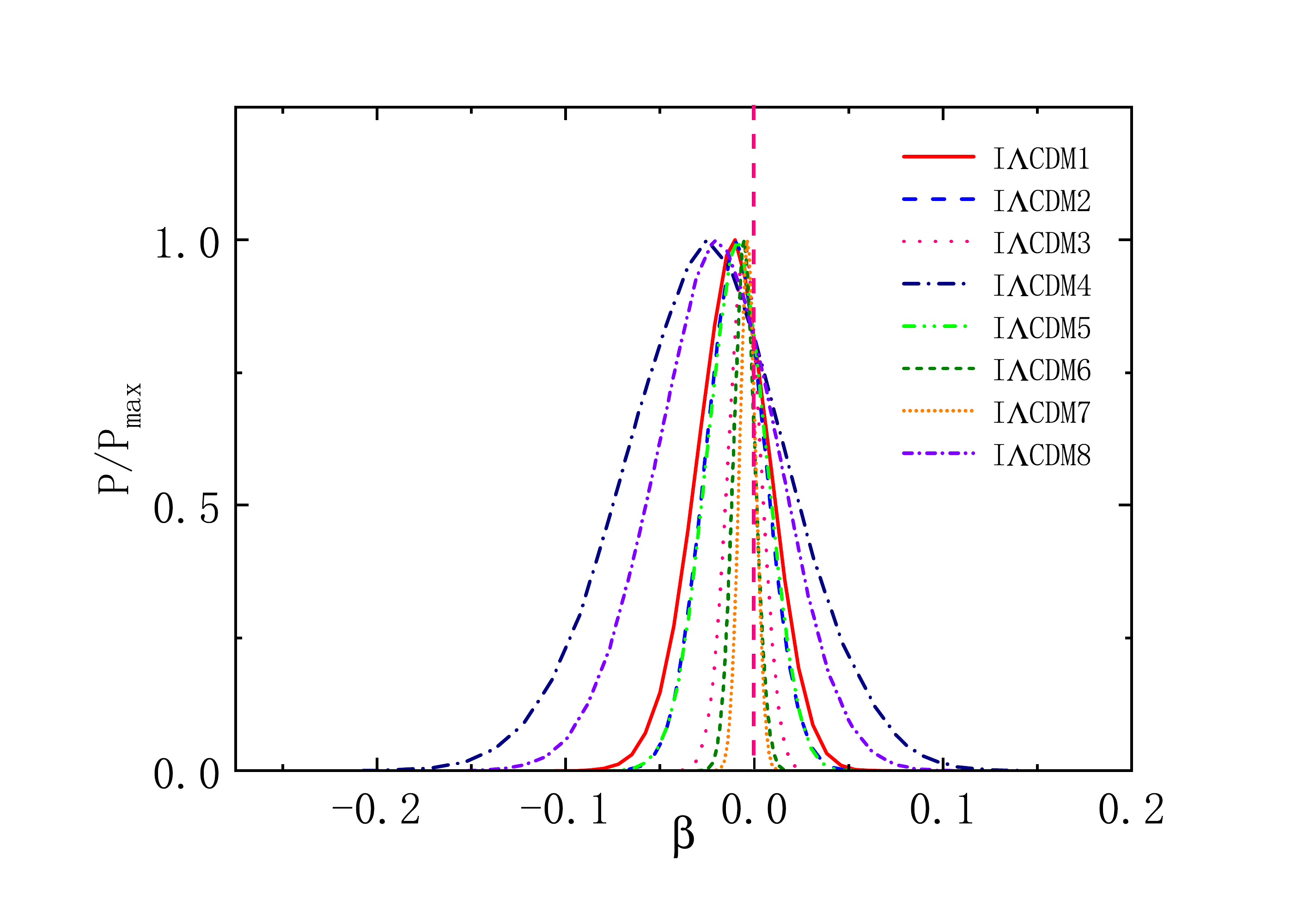}
\caption{\label{fig4}One-dimensional marginalized posterior distributions of the parameter $\beta$ for the I$\Lambda$CDM models, from the SN+CMB+BAO+$H_0$ data. The pink dashed line denotes the case of $\beta=0$.}
\end{figure*}

\begin{figure*}[!htp]
\includegraphics[width=7cm]{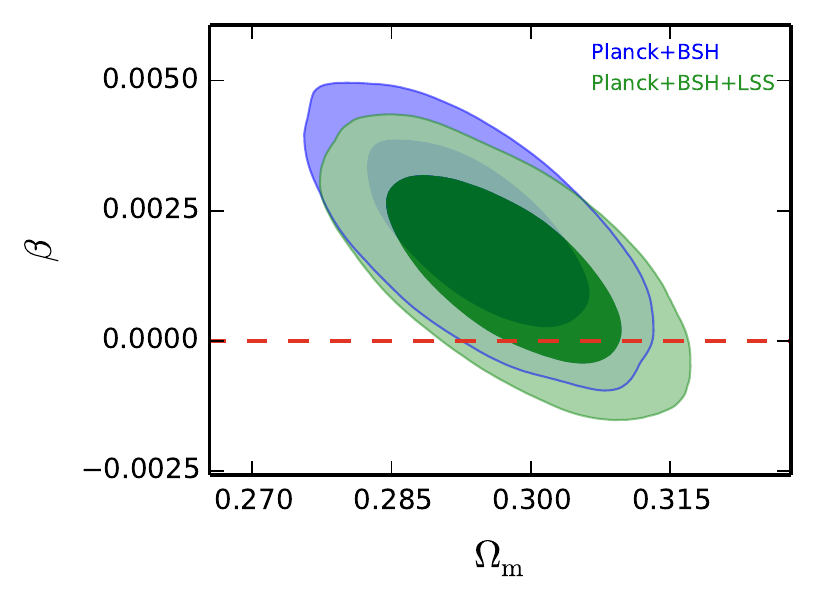}
\includegraphics[width=7cm]{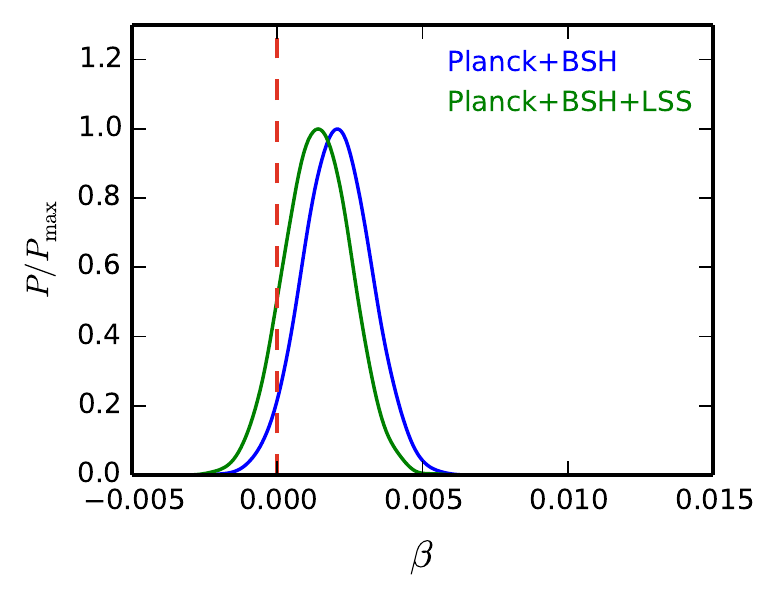}
\caption{\label{fig5}The two-dimensional marginalized contours (1$\sigma$ and 2$\sigma$) in the $\Omega_{\rm m}$--$\beta$ plane and the one-dimensional marginalized distributions of $\beta$ for the I$\Lambda$CDM model with $Q=-\beta H\rho_{\rm c}$ by using the Planck+BSH and Planck+BSH+LSS data. The red dashed line denotes the case of $\beta=0$. }
\end{figure*}

\begin{figure*}[!htp]
\includegraphics[width=7cm]{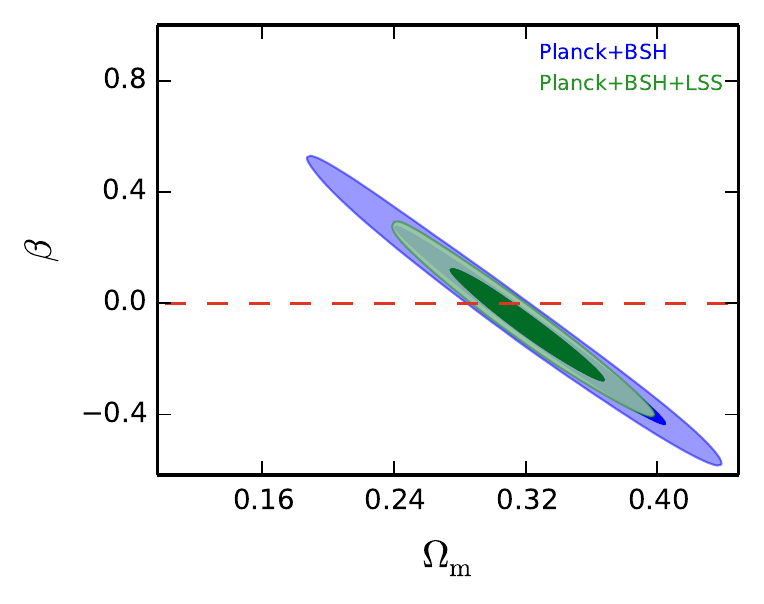}
\includegraphics[width=7cm]{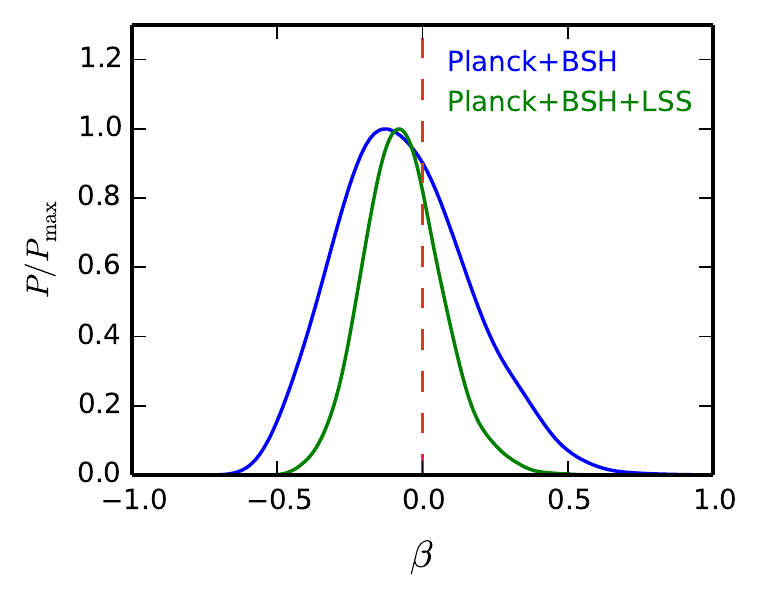}
\caption{\label{fig6}The two-dimensional marginalized contours (1$\sigma$ and 2$\sigma$) in the $\Omega_{\rm m}$--$\beta$ plane and the one-dimensional marginalized distributions of $\beta$ for the I$\Lambda$CDM model with $Q=-\beta H\rho_{\rm vac}$ by using the Planck+BSH and Planck+BSH+LSS data. The red dashed line denotes the case of $\beta=0$. }
\end{figure*}

In this section, we report the fitting results of the eight typical I$\Lambda$CDM models, and then discuss the implications of these results. We use the observational data combination SN+CMB+BAO+$H_0$ to constrain the cosmological parameters. The detailed fit values of cosmological parameters are presented in Tables~\ref{table1}--\ref{table3}.

In Table~\ref{table1}, we list the values of $\chi^2_{\rm min}$, $\Delta$AIC, and $\Delta$BIC for these models. Compared with the $\Lambda$CDM model, the $\chi^2_{\rm min}$ values of these I$\Lambda$CDM models are slightly smaller. But, when considering the factor of number of parameters, the $\Lambda$CDM model is still the best one. We choose the $\Lambda$CDM model as a reference model, so the values of $\Delta$AIC and $\Delta$BIC for it are both zero, and the I$\Lambda$CDM models have $\Delta {\rm AIC}\sim 1.2-1.7$ and $\Delta {\rm BIC} \sim 5.8-6.3$. In addition, we also find that the values of $\chi^2_{\rm min}$ for all the I$\Lambda$CDM models are almost equal (about 699), indicating that the current observational data almost equally favor the eight I$\Lambda$CDM models. From Fig.~\ref{fig1}, we can also clearly see that $\Delta$AIC (also $\Delta$BIC) of these I$\Lambda$CDM models are almost equal. A detailed comparison for them in the cosmological fit reveals that the best one is the I$\Lambda$CDM6 model and the next best one is the I$\Lambda$CDM7 model.

In this work, we study the I$\Lambda$CDM cosmology, and we have found that for this scenario the models with different forms of $Q$ are almost equally favored by the current observations. But for other types of interacting dark energy cosmology, the situation might be different. For example, the interacting holographic dark energy (IHDE) models with different forms of $Q$ have also been studied in the literature and a comparison has also been performed for them. In Ref.~\cite{Li:2017usw}, it was shown that all the IHDE models with $Q=\beta H_0\rho$ are equally favored by using the same data combination as the present work (SN+CMB+BAO+$H_0$). However, Ref.~\cite{Feng:2016djj} showed that in the IHDE cosmology with $Q=\beta H\rho$, the $Q=\beta H\frac{\rho_{\rm{vac}}\rho_{c}}{\rho_{\rm{vac}}+\rho_{\rm c}}$ model is most favored by the observational data, and the $Q=\beta H\rho_{c}$ model is not favored by the observational data, although the observational data are also the same as used in the present work.

%Compared with the I$\Lambda$CDM model, the $\chi^2_{\rm min}$ of IHDE model with different $Q$ ($Q=\beta H_0\rho$ and $Q=\beta H\rho$) are a little different. This indicates that the interaction forms of $Q$ in the IHDE models have a slightly greater impact on $\chi^2_{\rm min}$; but the interaction forms of $Q$ in the I$\Lambda$CDM models have a slightly smaller effect on $\chi^2_{\rm min}$.

The fitting results of the parameters of the $\Lambda$CDM model and these I$\Lambda$CDM models with $Q=\beta H_0\rho$ and $Q=\beta H\rho$ are listed in Tables~\ref{table2} and \ref{table3}. From the two tables, we find that the current observations slightly favor a negative coupling parameter $\beta$ (although $\beta=0$ is still within the 1$\sigma$ range).

\begin{table*}[!htp]\small
\setlength\tabcolsep{5pt}
\caption{Fitting results of the I$\Lambda$CDM models with $Q=\beta H\rho_{\rm vac}$ from different datasets. Best-fit values with $\pm1\sigma$ errors are presented. }
\label{table4}
\renewcommand{\arraystretch}{2}\centering
\begin{tabular}{lccccc}
\\
\hline
Parameter  & CMB & BAO& SN&$H_{0}$\\
  \hline
$\Omega_{\rm{m0}}$ & $0.5256^{+0.0791}_{-0.3107}$
                   & $0.4577^{+0.0691}_{-0.3326}$
                   & $0.3713^{+0.0539}_{-0.0957}$
                   & $0.8918^{+0.7060}_{-0.8806}$
                   \\

$\Omega_{\rm{b0}}$ & $0.0710^{+0.0072}_{-0.0340}$
                   & $0.1091^{+0.1216}_{-0.0991}$
                   & $0.2594^{+0.1521}_{-0.2494}$
                   & $0.5960^{+0.4040}_{-0.5859}$
                   \\

$\beta$            & $-0.5461^{+0.6280}_{-0.4539}$
                   & $0.9671^{+0.0328}_{-1.9671}$
                   & $0.7202^{+0.2798}_{-0.8374}$
                   & $0.2719^{+0.7281}_{-1.2719}$
                   \\

$h$                & $0.5603^{+0.2134}_{-0.0276}$
                   & $0.6532^{+0.3468}_{-0.2655}$
                   & $0.4881^{+0.5119}_{-0.1881}$
                   & $0.7060^{+0.0330}_{-0.0330}$
                   \\
\hline
\end{tabular}
%\caption{Fitting results of the models.}
\end{table*}

\begin{table*}[!htp]\small
\setlength\tabcolsep{5pt}
\caption{Fitting results of the I$\Lambda$CDM models with $Q=\beta H\rho_{\rm c}$ from different datasets. Best-fit values with $\pm1\sigma$ errors are presented. }
\label{table5}
\renewcommand{\arraystretch}{2}\centering
\begin{tabular}{lccccc}
\\
\hline
Parameter  & CMB & BAO& SN&$H_{0}$\\
  \hline
$\Omega_{\rm{m0}}$ & $0.3044^{+0.3188}_{-0.1188}$
                   & $0.4266^{+0.1192}_{-0.1617}$
                   & $0.3538^{+0.0458}_{-0.0729}$
                   & $0.8683^{+0.7307}_{-0.8574}$
                   \\

$\Omega_{\rm{b0}}$ & $0.0490^{+0.0010}_{-0.0263}$
                   & $0.0449^{+0.1254}_{-0.0349}$
                   & $0.0224^{+0.2829}_{-0.0124}$
                   & $0.5786^{+0.4214}_{-0.5684}$
                   \\

$\beta$            & $0.0051^{+0.0850}_{-0.0955}$
                   & $0.8252^{+0.1748}_{-0.9324}$
                   & $0.8651^{+0.1349}_{-1.18544}$
                   & $-0.4657^{+1.4657}_{-0.5343}$
                   \\

$h$                & $0.6742^{+0.3174}_{-0.0088}$
                   & $0.3804^{+0.6196}_{-0.0633}$
                   & $0.8290^{+0.1710}_{-0.5290}$
                   & $0.7060^{+0.0330}_{-0.0330}$
                   \\
\hline
\end{tabular}
%\caption{Fitting results of the models.}
\end{table*}

In Figs.~\ref{fig2} and \ref{fig3}, we show the two-dimensional posterior distribution contours (1$\sigma$ and 2$\sigma$) for these I$\Lambda$CDM models in the $\Omega_{\rm m0}$--$\beta$ and $\beta$--$h$ planes, respectively. To clearly display the results of different I$\Lambda$CDM models, we use the blue and pink contours to represent the $Q=\beta H_0\rho$ models and the $Q=\beta H\rho$ models, respectively.
From the two figures, we can clearly see that in all the I$\Lambda$CDM models $\beta=0$ is consistent with the current data inside the 1$\sigma$ range.
In addition, we also find that the constraints on $\beta$ for the $Q=\beta H\rho$ models are tighter than those for the $Q=\beta H_0\rho$ models.
From Fig.~\ref{fig2}, we can see that the constraints on $\Omega_{\rm m0}$ for the two classes of models ($Q=\beta H_0\rho$ and $Q=\beta H \rho$) are rather similar, while the constraints on $\beta$ for the $Q=\beta H \rho$ models are obviously tighter than those for the $Q=\beta H_{0} \rho$ models, for which the most prominent cases are given by the $Q\propto\rho_{\rm c}$ models, and next by the $Q\propto(\rho_{\rm vac}+\rho_{\rm c})$ models.
From Fig.~\ref{fig3}, we can see the situation for the constraints on $h$. For the $Q\propto\rho_{\rm vac}$ models and the $Q\propto\rho_{\rm vac}\rho_{\rm c}/(\rho_{\rm vac}+\rho_{\rm c})$ models, the constraints on $h$ for the two classes of models are similar. But for the $Q\propto\rho_{\rm c}$ models and the $Q\propto(\rho_{\rm vac}+\rho_{\rm c})$ models, the constraints on $h$ for the two classes of models are rather different; for this point, the $Q=\beta H_{0} \rho$ case is evidently better than the $Q=\beta H \rho$ case.
In short, from Figs.~\ref{fig2} and \ref{fig3}, we find that, for the constraints on the coupling parameter $\beta$, the $Q=\beta H \rho$ case is better than the $Q=\beta H_{0} \rho$ case.

In Fig.~\ref{fig4}, we show the one-dimensional marginalized posterior distributions of $\beta$ for the I$\Lambda$CDM models by using the SN+CMB+BAO+$H_0$ data. In this figure, the pink dashed line  denotes the case of $\beta=0$.
We can clearly see that for the constraints on the coupling parameter $\beta$, among these models the best one is the I$\Lambda$CDM7 model, and the next best ones are the I$\Lambda$CDM6 and I$\Lambda$CDM3 models; the worst one is the I$\Lambda$CDM4 model, and the next worst one is the I$\Lambda$CDM8 model. We also find that although a negative $\beta$ is slightly more favored, $\beta=0$ is still consistent with the data.

In this work, we use the joint SN+CMB+BAO+$H_0$ data to constrain the I$\Lambda$CDM models. It is however also important to look separately at constraints from each single dataset, in order to check the consistency of constraints from different datasets on the same model. We thus make such a consistency check. In Tables \ref{table4} and ~\ref{table5}, we show the fit results of the $Q=\beta H \rho_{\rm vac}$ and $Q=\beta H \rho_{\rm c}$ models by separately using the single dataset alone. We find that single dataset alone can only provide rather weak constraints for the cosmological parameters, and no evident inconsistency can be found in this check.

Note also that in this work we only consider the background constraints on the I$\Lambda$CDM cosmology, i.e., the cosmological perturbations are not considered in the calculations and observations from the growth of structures are not used in the constraints. It is of course meaningful to ask if the cosmological perturbations have important impacts on the I$\Lambda$CDM cosmology. However, it is well known that it is rather difficult to calculate the cosmological perturbations in a cosmology of dark energy interacting with dark matter, because in such a cosmology the cosmological perturbations will be divergent in a part of the parameter space, which ruins the whole calculation \cite{Zhao:2005vj,Valiviita:2008iv,He:2008si}. To overcome this instability difficulty, an effective theoretical framework based on an extended parametrized post-Friedmann (PPF) method has been established and developed by Yun-He Li, Jing-Fei Zhang, and Xin Zhang \cite{Li:2014cee,Li:2014eha,Zhang:2017ize}, by which the cosmological perturbations can be calculated safely in the whole parameter space of an interacting dark energy model and the observations of growth of structures can be used to constrain such a model.

In the following we will make some discussions on the issue concerning the cosmological perturbations by employing the extended PPF method. We do not wish to redo the analysis for all the eight models in this paper, and we actually only take two typical models as examples to elucidate the issue. Actually, in a recent paper by Rui-Yun Guo, Yun-He Li, Jing-Fei Zhang, and Xin Zhang \cite{Guo:2017hea}, the cases of $Q=-\beta H\rho_{\rm{c}}$ and $Q=-\beta H\rho_{\rm{vac}}$ have been analyzed by using the observations of both the expansion history and the growth of structures. Note here that because in Ref.~\cite{Guo:2017hea} the sign of $\beta$ is opposite with ours in this paper, we use the forms of $Q=-\beta H\rho_{\rm{c}}$ and $Q=-\beta H\rho_{\rm{vac}}$ to make discussions in this place. Thus, here, $\beta>0$ denotes dark matter decaying into vacuum energy, and $\beta<0$ denotes vacuum energy decaying into dark matter. In Ref.~\cite{Guo:2017hea}, two data combinations, Planck+BAO+SN+$H_0$ (Planck+BSH) and Planck+BAO+SN+$H_0$+RSD+WL (Planck+BSH+LSS), are used to constrain the models, where ``BSH'' is an abbreviation to denote the combination of BAO+SN+$H_0$, and ``LSS'' is an abbreviation of large-scale structure to denote the combination of RSD+WL (redshift space distortions and weak lensing). Here, the Planck 2015 full data of CMB power spectra (TT, EE, TE+lowP) are used, and the BAO data are from Refs.~\cite{Beutler:2011hx,Ross:2014qpa,Gil-Marin:2016wya}, the SN data are from JLA compilation \cite{Betoule:2014frx}, the $H_0$ measurement is from the 2016 measurement of Riess et al. \cite{Riess:2016jrr}, the RSD data are from the LOWZ sample of $z_{\rm eff}=0.32$ and the CMASS sample of $z_{\rm eff}=0.57$ \cite{Gil-Marin:2016wya}, and the WL data are from CFHTLenS \cite{Heymans:2013fya}. The constraint results (1$\sigma$ and 2$\sigma$ contours in the $\Omega_{\rm m0}$--$\beta$ plane and one-dimensional posterior distributions of $\beta$) are shown in Figs. \ref{fig5} and \ref{fig6}, which are directly plotted by using the likelihood chains obtained in Ref.~\cite{Guo:2017hea}. From these figures, we find that although the LSS data can help shrink the parameter space, in particular for the $Q=-\beta H\rho_{\rm{vac}}$ case, the current constraints from the cosmological perturbations and the LSS datasets are still not decisive in discriminating such models from the uncoupled $\Lambda$CDM cosmology.

\section{Conclusion}\label{sec4}

In this paper, we have investigated the models of vacuum energy interacting with cold dark matter from the perspective of observational constraints. We consider eight typical I$\Lambda$CDM models with the interaction terms $Q=\beta H_{0}\rho_{\rm{vac}}$, $Q=\beta H_{0}\rho_{\rm c}$, $Q=\beta H_{0}(\rho_{\rm{vac}}+\rho_{\rm c})$, $Q=\beta H_{0}\frac{\rho_{\rm vac}\rho_{\rm c}}{\rho_{\rm vac}+\rho_{\rm c}}$, $Q=\beta H\rho_{\rm{vac}}$, $Q=\beta H\rho_{\rm c}$, $Q=\beta H(\rho_{\rm{vac}}+\rho_{\rm c})$, and $Q=\beta H\frac{\rho_{\rm vac}\rho_{\rm c}}{\rho_{\rm vac}+\rho_{\rm c}}$; the former four interaction terms include the Hubble constant $H_{0}$, while the latter four include the Hubble parameter $H$. We place constraints on these models by using the current observational data, including the JLA compilation of type Ia supernovae, the CMB distance priors of Planck 2015, the BAO measurements, and the $H_{0}$ direct measurement.

By comparing $\chi^2_{\rm min}$ values and information criteria ($\Delta \rm{AIC}$ and $\Delta \rm{BIC}$) of the $\Lambda$CDM model and the eight I$\Lambda$CDM models, we find that the observational data sets (SN+CMB+BAO+$H_0$) almost equally favor these I$\Lambda$CDM models, and the $\chi^2_{\rm min}$ values of these I$\Lambda$CDM models are all around 699. We also find that in all these I$\Lambda$CDM models the coupling parameter $\beta=0$ is consistent with the current observational data within 1$\sigma$ range, indicating that the standard $\Lambda$CDM model is still well consistent with the current data, and a nonzero coupling is not detected. In addition, in this work we find that for the constraints on the coupling parameter $\beta$, the $Q=\beta H \rho$ case is better than the $Q=\beta H_{0} \rho$ case.
%But here we also wish to emphasize that in this study we did not consider the cosmological perturbations and we did not use the full CMB data of Planck, which would affect the results and the conclusion to some extent.

%When using the Planck+BSH to constrain the I$\Lambda$CDM model in the case of considering the cosmological perturbations, for the I$\Lambda$CDM model with $Q=\beta H\rho_{\rm{c}}$, $\beta>0$ is favored by observational data, for the I$\Lambda$CDM model with $Q=\beta H\rho_{\rm{vac}}$, $\beta<0$ is slightly favored by observational data, this implies that the cosmological perturbations and the observational data have some impacts on the determination of the coupling constant $\beta$; when adding the large-scale structure observational data RSD and WL, it can make the constraint of $\beta$ more tighter, but it cannot change the directions of energy flow for both $Q=\beta H\rho_{\rm{c}}$ and $Q=\beta H\rho_{\rm{vac}}$ models.

\acknowledgments
This work was supported by the National Natural Science Foundation of China (Grant Nos.~11875102,
11835009, 11522540, and 11690021), the National Program for Support of Top-Notch Young Professionals,
and Doctoral Research Project of Shenyang Normal University (Grant Nos.~BS201844 and BS201702).

\end{document}